\documentstyle[12pt,epsf]{article}

\newcommand{\vspu}{\vspace*{5mm}}

\newcommand\be{\begin{equation}}
\newcommand\ee{\end{equation}}
\newcommand\bea{\begin{eqnarray}}
\newcommand\eea{\end{eqnarray}}

\makeatletter \@addtoreset{equation}{section} \makeatother

\begin{document}
\begin{center}
{{\large\bf Reexamined radiative Decay \protect$
B_{q}^{*}\rightarrow B_{q}\gamma \protect$ in Light Cone QCD }
\vspu \\
\ \\
Zuo-Hong Li\protect$ ^{a,b,c,d,}\protect $\footnote{Email: lizh@ytu.edu.cn}, Xiang-Yao
Wu\protect$ ^{c}\protect $, and Tao Huang\protect$ ^{a,c}\protect $
\\
\vspu \ \\
\footnotesize{\small{ a.CCAST (World Laboratory), P.O.Box 8730, Beijing 100080, China}
\\ \small{b.Department of Physics, Yantai University, Yantai 264005,
China\footnote{Mailing address}}\\ \small{c. Institute of High Energy Physics, P.O.Box
918(4), Beijing 100039, China}\\ \small {d. Department of Physics, Peking University,
Beijing 100871, China }
\\
}}
\end{center}
\vspu
\begin{center}{\bf Abstract}
\end{center}
\normalsize

The radiative decay $ B_{q}^{*}\rightarrow B_{q}\gamma( q=u,d\,
or\, s)  $ is reexamined with a modified light-cone QCD sum rule
method, in which adequate chiral operators are chosen as the
interpolating fields in the correlators used for a sum rule
estimate of the relevant coupling $ g_{B_{q}^{*}B_{q}\gamma } $.
The resulting sum rules not only show the physical picture
consistent with the underlying physics in $ B_{q}^{*}\rightarrow
B_{q}\gamma $ but also avoid the pollution by the nonlocal matrix
element $ \left\langle \gamma \left( q\right) \left|
\overline{q}\left( x\right) \gamma _{\mu }\gamma _{5}q\left(
0\right) \right| 0\right\rangle$, which starts with twist-3 and
thus may bring a large uncertainty to the sum rule predictions.
Also, a comparison is made with the previous results from
light-cone QCD sum rules and chiral perturbation theory.

\begin{verse}

\end{verse}
\newpage
\baselineskip 18pt
\begin{center}{\large \bf{1. INTRODUCTION}}\end{center}\indent

The physics of exclusive $ B $ decays is particularly interesting
for a detailed test of the standard model (SM) and a careful
search for potential signature of new physics. With a large amount
of available data from the BaBar and Belle, it is believed that a
considerable progress will be made in the area in the near future.
A good many theoretical works contribute to this subject. A recent
progress deserving mention is the presentation of the QCD
factorization formula for $B\to \pi\pi$, $\pi K$ and $\pi
D$\cite{1}. Nevertheless, to make a reliable estimate of physical
amplitudes demands that we calculate precisely the relevant
hadronic matrix elements. It is practically beyond our present
ability to do such a calculation from first principle. At present,
most of efforts have been devoted to looking for phenomenological
approaches to nonperturbative QCD dynamics. Among the most
successful models is QCD sum rule method \cite{2}. As it is often
case, however, this approach has a restricted application to
heavy-to-light transitions, failing to give a viable behavior of
the form factors with the heavy quark mass at small momentum
transfer. This originates technically from a cut-off of vacuum
condensate terms in the short distance expansion where only the
operators of lower dimension remain, and physically from neglect
of the finite correlations between the quarks or the gluons in the
physical vacuum. The case can improve by using the operator
product expansion (OPE) around the light cone $ x^{2}\approx 0 $
instead of at the short-distance $ x\approx 0 $. The calculational
framework established in such a way is termed light-cone QCD sum
rule {[}3,4{]}, which combines the well-developed description of
exclusive process in terms of perturbative QCD factorization
formula with traditional QCD sum rules. It has found wide
applications {[}3-11{]} in the literature, especially applying to
a study on heavy-to-light transitions, and the obtained
predictions are encouraging. However, new nonperturbative inputs,
the light-cone wavefunctions of light mesons classified by twists,
are involved in light-cone sum rules. A question we must answer is
whether these wavefunctions correctly reflect the underlying
nonperturbative dynamics, in other word, whether they are
phenomenologically better known. To date, only the components of
leading twist have undergone a systematic investigation {[}6{]}
for the distribution amplitudes of the nonsinglet, while the high
twist wavefunctions, some of which play a role as important as the
twist-2 ones in most cases, require a further examination or
refinement for improving the accuracy of numerical results. To
prevent the resulting sum rules from suffering the possibly
sizable contamination by the twist-3 components, an effective
prescription \cite{3,11} has been suggested to estimate
heavy-to-light form factors. The trick, to be specific, is to
choose an adequate chiral current correlator, as allowed by the
sum rules. As a result, not only the resulting new sum rules work
effectively but also some of the twist-3 components, which make
predominant contributions with respect to the other twist-3
wavefunctions in conventional light-cone QCD sum rule
calculations, vanish automatically. From the theoretical
viewpoint, this procedure offers a certain modification of the
existing light-cone QCD sum rule approach to heavy-to-light
transitions, making the nonperturbative effects well controlled.

Also, among the important exclusive processes are the $ B $ decays
with a photon emission. They have in fact received a variety of
model investigations. It is in {[}4{]} that the first application
was presented of light-cone QCD sum rule technique to this type of
processes, where the radiative decay $ \Sigma \rightarrow P\gamma
$ is discussed in detail. Later on, the same approach was applied
to study $ B\rightarrow \rho \gamma  $ {[}7,8{]}, $ B\rightarrow
l\nu _{l}\gamma  $ {[}9{]} and $ B_{q}^{*}\rightarrow B_{q}\gamma
\left( q=u,\ d\, or\, s\right)  $ {[}10{]}. However, the problems
with $ B_{q}^{*}\rightarrow B_{q}\gamma  $ have been at issue. The
yielded decay widths in different approaches disagree in numerical
results and, to some extent, also in order of magnitude. For
instance, light-cone QCD sum rules {[}10{]} predict $ \Gamma
\left( B_{u}^{*}\rightarrow B_{u}\gamma \right) =0.63\, KeV $ and
$ \Gamma \left( B_{d}^{*}\rightarrow B_{d}\gamma \right) =0.16\,
KeV $ while chiral perturbation theory {[}12{]} gives $ \Gamma
\left( B_{u}^{*}\rightarrow B_{u}\gamma \right) =0.14\, KeV $ and
$ \Gamma \left( B_{d}^{*}\rightarrow B_{d}\gamma \right) =0.09\,
KeV $. A definite conclusion depends strongly on our ability to
calculate the relevant $B^{\ast} _qB_q\gamma$ coupling constant $
g_{B_{q}^{*}B_{q}\gamma }$, which is intrinsically a
nonperturbative quantity and is defined as
\begin{equation}
\left\langle \gamma \left( q,\eta \right) B_{q}^{*}\left(
p,e\right) \mid B_{q}\left( p+q\right) \right\rangle
=g_{B_{q}^{*}B_{q}\gamma }\epsilon _{\alpha \beta \rho \sigma
}p^{\alpha }q^{\beta }\eta ^{\rho }e_{\lambda }^{\sigma }.
\end{equation}
where $\eta_{\mu}\ (e_{\mu}), q\ (p)$ are the polarization vector
and the momentum of the photon(the $B^*$ meson) respectively.

In the paper we would reexamine the radiative decay $
B_{q}^{*}\rightarrow B_{q}\gamma $ by performing a calculation on
the coupling $ g_{B_{q}^{*}B_{q}\gamma } $ with an improved
light-cone QCD sum rule approach, in which the appropriate chiral
currents operators act as the interpolating fields. Our findings,
are intriguing: The resulting sum rules may not only be of a clear
physical meaning but also avoid the pollution by the nonlocal
matrix element$ \left\langle \gamma \left( q\right) \right|
\overline{q}\left( x\right) \gamma _{\mu }\gamma _{5}q\left(
0\right) \left| 0\right\rangle$ starting with twist-3, which is
known very poorly and thus may bring a large uncertainty to the
sum rule results.

This presentation is organized as follows. In the following
section we propose our improved approach. Then a detailed light
cone QCD calculation is preformed of the relevant correlator in
Section 3. Section 4 is devoted to a numerical discussion on the
sum rule results for $g_{B_{q}^{*}B_{q}\gamma }$ and then an
estimate of $\Gamma \left( B_{q}^{*}\rightarrow B_{q}\gamma
\right)$. We give a simple summary in the last Section.

\begin{center}{\large \bf 2. CHIRAL CURRENT CORRELATOR}\end{center}

At the quark level, the radiative transition $
B_{q}^{*}\rightarrow B_{q}\gamma  $ is dominated by the magnetic
moment interaction between the light quark and electromagnetic
field due to small recoil of the decaying meson. This fact
reveals, from the viewpoint of QCD, that the process in question
belongs in the long distance type. It is well known that the
current operators used as interpolating fields are by no means
uniquely determined in QCD sum rules, and it is allowed to take
different correlators providing that the validity of the resulting
sum rule remains. This offers a possibility to explore the
radiative decay $B_{q}^{*}\rightarrow B_{q}\gamma $ in a way
consistent with the physical picture. In order to obtain such a
desired sum rule result for $ g_{B_{q}^{*}B_{q}\gamma }$ it is in
order to adopt the following correlator,
\begin{eqnarray}
F_{\mu }\left( p^{2},\left( p+q\right)^2 \right)  & = & i\int d^{4}xe^{ipx}\left\langle \gamma \left( q,\eta \right) \left| TJ_{\mu }^{V-A}\left( x\right) J^{S+P}\left( 0\right) \right| 0\right\rangle \nonumber \\
 & = & \epsilon _{\mu \nu \alpha \beta }p^{\nu }q^{\alpha }\eta ^{*\beta }F\left( p^{2},\left( p+q\right)^{2} \right) +Vector\, Part,
\end{eqnarray}
where the chiral current operators $ J_{\mu
}^{V-A}=\overline{q}\gamma _{\mu }\left( 1-\gamma _{5}\right) b $
and $ J_{\mu }^{S+P}=\overline{b}i\left( 1+\gamma _{5}\right) q $
are chosen as the interpolating fields. For the invariant function
$F\left( p^{2},\left( p+q\right)^{2} \right) $, remarkably, the
resulting hadronic representation $F^H\left( p^{2},\left(
p+q\right)^{2} \right)$ depends not only on resonances of $
J^{P}=0^{-},1^{-} $ but also on those of $ J^{P}=0^{+},\ 1^{+}. $
However, we can safely isolate the pole contribution of $ B_{q} $
and $ B_{q}^{*} $ , due to a sizable mass difference between the
lowest negative and positive parity states, and express the high
state contribution starting from the threshold parameters $ s_{0}
$ and $ s^{'}_{0} $, which are set around the squared masses of
the lowest $ 1^{+} $ and $ 0^{-} $ states respectively, in a form
of dispersion integral. The result is
\begin{eqnarray}
F^{H}\left( p^{2},\! (p+q\right)^2  & = & \frac{m_{B_{q}}^{2}m_{B_{q}}^{*}}{m_{b}+m_{q}}\frac{f_{B_{q}}f_{B_{q}^{*}}g_{B_{q}^{*}B_{q}\gamma }}{\left( p^{2}-m_{B^{*}_{q}}^{2}\right) \left[ \left( p+q\right) ^{2}-m_{B_{q}}^{2}\right] }\nonumber \label{mathed:} \\
 & + & \int \int ds_{1}ds_{2}\frac{\rho ^{H}\left( s_{1},s_{2}\right) }{\left( p^{2}-s_{1}\right) \left[ \left( p+q\right) ^{2}-m^{2}_{B_{q}}\right] },
\end{eqnarray}
with the decay constants defined as
\begin{equation}
\left\langle B\right| \overline{b}i\gamma _{5}q\left| 0\right\rangle =\frac{m_{B_{q}}^{2}}{m_{b}+m_{q}}f_{B_{q}},
\end{equation}
\begin{equation}
\left\langle 0\left| \overline{q}\gamma _{\mu }b\right| B_{q}^{*}\right\rangle
=m_{B_{q}}f_{B^{*}_{q}}e_{\mu }^{\left( \lambda \right) }.
\end{equation}
After the Borel improvement $ P^{2}\rightarrow M_{1}^{2},\, \left( p+q\right)
^{2}\rightarrow M_{2}^{2}, $ Eq. (3) is of the following form:
\begin{equation}
F^{H}\left( M_{1}^{2},M_{2}^{2}\right) =\frac{1}{M_{1}^{2}M_{2}^{2}}\left( \frac{m_{B_{q}}
^{2}m_{B_{q}}^{*}}{m_{b}+m_{q}}f_{B_{q}}f_{B^{*}_{q}}g_{B^{*}_{q}B_{q}\gamma }e^{-\frac{m^{2}
_{B^{*}_{q}}}{M_{1}^{2}}-\frac{m^{2}_{B_{q}}}{M_{2}^{2}}}+\int \int ds_{1}ds_{2}\rho ^{H}
\left( s_{1},s_{2}\right) e^{-\frac{s_{1}}{M_{1}^{2}}-\frac{s_{2}}{M_{2}^{2}}}\right) .
\end{equation}

\begin{center}{\large \bf 3. LIGHT CONE OPE}\end{center}

In this section, we will make use of the light cone OPE and give a
detailed derivation of the theoretical expression for the
correlator in Eq.(2). Schematically, the main contributions in the
OPE come from the four Feynman diagrams shown in Fig. 1. All these
diagrams can be calculated one by one for large space-like
momentum regions $ p^{2}\ll 0 $ and $ \left( p+q\right) ^{2}\ll 0
$, where the $ b $ quark is far off shell so that the OPE goes
effectively.
\begin{center}{\bf (1). SHORT DISTANCE CONTRIBUTION }\end{center}

The short distance contribution, as will be shown, is less
important from Fig.1 (a) and (b) but its derivation turns out to
be more complicated with respect to the long distance case. We
have the UV-convergent loop integrals to deal with:
\begin{eqnarray}
I=\int d^4k\frac{1}{ [(p+k)^2-m_b^2] [(k+p+q)^2-m_b^2](k^2-m_q^2)}, \nonumber\\
I^{\mu}=\int d^4k\frac{k^{\mu}}{[(p+k)^2-m_b^2]
[(k+p+q)^2-m_b^2](k^2-m_q^2)},
\end{eqnarray}
for Fig.1(a) and the similar ones for Fig.1(b). For the
convenience of making the Borel transformation, we use the forms
of Feynman parametrization for them. After doing this, performing
the loop integrals we have the invariant function $
F(p^{2},(p+q)^2) $,
\begin{equation}
F^{\left( a\right) }(p^2,(p+q)^2)=\frac{3Q_{b}}{2\pi ^{2}}\int
_{0}^{1}dx\int
_{0}^{1}dy\frac{m_{q}x\overline{x}}{m_{b}^{2}x+m_{q}^{2}\overline{x}-p^{2}x\overline{x}\,
\overline{y}-\left( p+q\right) ^{2}x\overline{x}y}
\end{equation}
for Fig.1 (a) and
\begin{equation}
F^{\left( b\right) }(p^{2},(p+q)^2)=\frac{3Q_{q}}{2\pi ^{2}}\int
_{0}^{1}dx\int
_{0}^{1}dy\frac{m_{q}x^{2}}{m_{b}^{2}\overline{x}+m_{q}^{2}x-p^{2}x\overline{x}\,
\overline{y}-\left( p+q\right) ^{2}x\overline{x}y}
\end{equation}
for Fig.1 (b), with $ \overline{x}=1-x $ and $ \overline{y}=1-y $.
The total perturbative contribution can be written as
\begin{equation}
F^{\left( a+b\right) }(p^{2},(p+q)^2)=\frac{3m_{q}}{2\pi ^{2}}\int
_{0}^{1}dx\int _{0}^{1}dy\frac{\left(
Q_{q}x+Q_{b}\overline{x}\right)
x}{m_{b}^{2}\overline{x}+m_{q}^{2}x-p^{2}x\overline{x}\,
\overline{y}-\left( p+q\right) ^{2}x\overline{x}y}.
\end{equation}
To express it in a form of dispersion integral we use the exponential form
\begin{equation}
F^{\left( a+b\right) }(p^{2},(p+q)^2)=\frac{3m_{q}}{2\pi ^{2}}\int
_{0}^{1}dx\int _{0}^{1}dy\int ^{\infty }_{0}d\alpha x\left(
Q_{q}x+Q_{b}\overline{x}\right) e^{-\alpha \left(
m_{b}^{2}\overline{x}+m_{q}^{2}x+Q_{1}^{2}x\overline{x}\,
\overline{y}+Q_{2}^{2}x\overline{x}y\right) },
\end{equation}
where $ Q_{1}^{2}=-p^{2} $ and $ Q_{2}^{2}=-\left( p+q\right) ^{2}. $ Following the
method proposed in {[}13{]}, we arrive immediately at the perturbative spectral
density $ \rho ^{\left( a+b\right) }\left( s_{1},s_{2}\right)  $,
\begin{equation}
\rho ^{\left( a+b\right) }\left( s_{1},s_{2}\right) =\frac{3m_{q}}{2\pi ^{2}}\int _{x_{1}}^{x_{2}}\frac{dx}{\overline{x}}\left( m_{q}x+m_{b}\overline{x}\right) \delta \left( s_{1}-s_{2}\right) \Theta \left( s_{1}-\left( m_{b}+m_{q}\right) ^{2}\right) \Theta \left( s_{2}-\left( m_{b}+m_{q}\right) ^{2}\right)
\end{equation}
with
\begin{equation}
x_{1}=\left( s_{1}+m_{b}^{2}-m_{q}^{2}-\sqrt{s_{1}^{2}-2\left( m_{b}^{2}+m_{q}^{2}\right) s_{1}+\left( m_{b}^{2}-m_{q}^{2}\right) ^{2}}\right) /2s_{1},
\end{equation}

\begin{equation}
x_{2}=\left( s_{1}+m_{b}^{2}-m_{q}^{2}+\sqrt{s_{1}^{2}-2\left( m_{b}^{2}+m_{q}^{2}\right) s_{1}+\left( m_{b}^{2}-m_{q}^{2}\right) ^{2}}\right) /2s_{1}.
\end{equation}
Furthermore, completing the integral over $ x $ yields
\begin{eqnarray}
\rho ^{\left( a+b\right) }\left( s_{1},s_{2}\right)  & = & \frac{3m_{q}}{2\pi ^{2}}\delta \left( s_{1}-s_{2}\right) \Theta \left( s_{1}-\left( m_{b}+m_{q}\right) ^{2}\right) \Theta \left( s_{2}-\left( m_{b}+m_{q}\right) ^{2}\right) \nonumber \\
 & \times  & \left[ \left( Q_{b}-Q_{q}\right) \lambda \left( 1,A,B\right) +Q_{q}\ln \frac{1-A+B+\lambda \left( 1,A,B\right) }{1-A+B-\lambda \left( 1,A,B\right) }\right] ,
\end{eqnarray}
with $ A=m_{b}^{2}/s_{1} $, $ B=m_{q}^{2}/s_{2} $ and $ \lambda \left( 1,A,B\right) $
defined as
\begin{equation}
\lambda \left( 1,A,B\right) =\sqrt{1+A^{2}+B^{2}-2A-2B-2AB}.
\end{equation}
Putting everything together, we have the Borel improved result
\begin{eqnarray}
\overline{F}^{\left( a+b\right) }\left( M_{1}^{2},M_{2}^{2}\right)  & = & \frac{3m_{q}}{2\pi ^{2}M_{1}^{2}M_{2}^{2}}\int ds\Theta \left( s-\left( m_{b}+m_{q}\right) ^{2}\right) e^{-\frac{s}{M^{2}}}\nonumber \\
 & \times  & \left[ \left( Q_{b}-Q_{q}\right) \lambda \left( 1,A,B\right) +Q_{q}\ln \frac{1-A+B+\lambda \left( 1,A,B\right) }{1-A+B-\lambda \left( 1,A,B\right) }\right] ,
\end{eqnarray}
with $ M^{2}=M_{1}^{2}M_{1}^{2}/(M_{1}^{2}+M_{1}^{2}). $ It is explicitly shown that
the short distance contribution is strongly suppressed from the perturbative hard
quarks owing to smallness of the light quark mass $ m_{q} $ , just as one may expect
on the basis of a naive analysis.
\begin{center}{\bf (2). LONG DISTANCE CONTRIBUTION}\end{center}

Fig.1 (c) and (d) denote the long distance effects induced by the
light quark propagating in vacuum and interaction of the light
quark in vacuum with the external electromagnetic field,
respectively. In the light cone OPE, these soft quarks in vacuum
are considered to be in a state of finite correlation, namely they
are of a non-vanishing momentum {[}14{]}, which stems from the
high dimension operators neglected in conventional QCD sum rule
calculations. We might describe the effect by introducing the
concept of nonlocal quark condensate {[}14{]}.

In the first place, we consider the contribution of Fig.1 (c). In
this case the nonlocal quark condensate $ \left\langle
\overline{q}(x)q(0)\right\rangle  $appears in the light cone OPE.
To model its behavior near the light-cone $ x^{2}=0 $, a pragmatic
strategy is to assume it to obey a distribution of the Gaussian
type {[}14{]}
\begin{equation}
\left\langle \overline{q}(x)q(0)\right\rangle =\left\langle \overline{q}q\right\rangle e^{-\frac{x^{2}_{E}}{4\rho }},
\end{equation}
with $ x_{E}^{2}=-x^{2}. $ If comparing Eq.(18) with the short
distance expansion

\begin{equation}
\left\langle \overline{q}(x)q(0)\right\rangle =\left\langle
\overline{q}q\right\rangle -\frac{m_{0}^{2}}{16}\left\langle
\overline{q}q\right\rangle x^{2}_{E}+\cdot \cdot \cdot
\end{equation}
with $ m^{2}_{0}=0.8\ GeV^{2}, $ the parameter $ \rho  $ is found to be $ 5\ GeV^{-2}
$. Introducing the parameters $ \alpha  $, $ \beta  $ and $ \gamma =\alpha +\beta
+\rho  $ and then performing the cumbersome loop-integral, we obtain the following
result for the contribution of the nonlocal quark condensate,

\begin{equation}
F^{(c)}(p^{2},(p+q)^{2})=-2\rho ^{3}Q_{b}\left\langle \overline{q}q\right\rangle \int _{0}^{\infty }d\alpha d\beta \frac{1}{\gamma ^{3}}e^{-\frac{\alpha \rho }{\gamma }P_{E}^{2}-\frac{\beta \rho }{\gamma }Q^{2}_{E}-(\alpha +\beta )m_{b}^{2}},
\end{equation}
where $ P^{2}_{E}=-p^{2},Q_{E}^{2}=-(p+q)^{2}. $ Furthermore, it can be converted into
the Borel transformed form
\begin{equation}
\overline{F}^{(c)}(M_{1}^{2},M_{2}^{2})=-2Q_{b}\left\langle \overline{q}q\right\rangle \frac{1}{M_{1}^{2}M_{2}^{2}}e^{-\rho m^{2}_{b}\left( \frac{M_{1}^{2}+M_{2}^{2}}{\rho M_{1}^{2}M_{2}^{2}-M_{1}^{2}-M_{2}^{2}}\right) }\Theta \left( \rho -\frac{1}{M_{1}^{2}}-\frac{1}{M_{2}^{2}}\right) .
\end{equation}
The $ \Theta  $ function restricts the Borel parameters $
M_{1}^{2} $ and $ M_{2}^{2} $ to the range $
\frac{1}{M_{1}^{2}}+\frac{1}{M_{2}^{2}}<\rho  $. We have already
checked that as $ \rho \rightarrow \infty  $ the above result
returns precisely to what it should be in the case in which only
the leading term remains in (19).

A subtle approach has been suggested to deal with that type of
dynamics in Fig.1 (d). The idea is to introduce so called photonic
light-cone wavefunctions to parametrize the long distance
dynamics. This can be explained specifically as follows: On
contracting the $b$ quark operators, one has the matrix element
$T\sim iQ_q\eta_{\rho}^{(\lambda)}\int dz e^{iq\cdot z}\langle 0|
J_{em}^{\rho}(z){\cal{O}}(x,0)| 0\rangle$, where
$J_{em}^{\rho}(z)$ is the electromagnetic current operator of a
light quark $q$ with the charge $Q_q$ and ${\cal{O}}(x,0)$
expresses a certain nonlocal operator. If the underlying light
quarks are hard we can perform the standard OPE. This is typically
the case of Fig.1 (b). For the soft light quarks we leave the
light quark operators without contraction and treat
$J_{em}^{\rho}(z)$ as the interpolating field of a photon state
with the momentum $q$. In such a way we can define a matrix
element of the nonlocal operator ${\cal{O}}(x,0)$ between the
vacuum and the photon state as
\begin{equation}
\langle \gamma(q)|{\cal{O}}(x,0)|
0\rangle=iQ_q\eta_{\rho}^{(\lambda)}\int dz\ e^{iq\cdot
z}\langle0\mid J_{em}^{\rho}(z){\cal{O}}(x,0)\mid 0\rangle,
\end{equation}
and further parametrize it via a series of photonic light cone
wavefunctions, which is equivalent to a summation over all the
relevant condensate terms in the case of the short distance OPE,
as in the applications of light cone QCD sum rules to $B\to \pi$
transitions. With such a trick, a straightforward calculation of
Fig.1 (d) yields
\begin{equation}
F_{\mu }^{(d)}(p^{2},(p+q)^{2})=\frac{2i}{\left( 2\pi \right) ^{4}}\int d^{4}x\int d^{4}ke^{i\left( p-k\right) \cdot x}\left\langle \gamma \left( q\right) \left| \overline{q}\left( x\right) \gamma _{\mu }\gamma _{\nu }\left( 1+\gamma _{5}\right) q\left( 0\right) \right| 0\right\rangle \frac{k^{\nu }}{m_{b}^{2}-k^{2}}.
\end{equation}
Using the $ \gamma  $ algebraic relations
\begin{equation}
\gamma _{\mu }\gamma _{\nu }=-i\sigma _{\mu \nu }+g_{\mu \nu },
\end{equation}

\begin{equation}
\gamma _{\mu }\gamma _{\nu }\gamma _{5}=-\frac{1}{2}\epsilon _{\mu \nu \alpha \beta }\sigma ^{\alpha \beta }+g_{\mu \nu }\gamma _{5},
\end{equation}
and considering the identities
\begin{eqnarray}
\left\langle \gamma \left( q\right) \left|
\overline{q}\left(x\right) \gamma _{5}q\left( 0\right)
\right|0\right\rangle &=&0,\\
\langle\gamma(q)|\bar{q}(x)q(0)|0\rangle &=& 0,
\end{eqnarray}
we have
\begin{eqnarray}
F_{\mu }^{(d)}(p^{2},(p+q)^{2}) & = & -\frac{2i}{\left( 2\pi
\right) ^{4}}\int d^{4}x\int d^{4}ke^{i\left( p-k\right) \cdot
x}\frac{k^{\nu }}{m_{b}^{2}-k^{2}}\left[i\langle \gamma \left(
q\right) \left| \overline{q}\left( x\right)
\sigma _{\mu \nu }q\left( 0\right) \right|0\rangle\nonumber\right. \\
&+&\left.\frac{1}{2}\epsilon _{\mu \nu \alpha
\beta}\langle\gamma\left( q\right)| \overline{q}\left(x\right)
\sigma^{\alpha \beta}q\left(0\right)|0\rangle\right].
\end{eqnarray}
An important observation is, as an appealing feature of our approach, that all the
long distance dynamics entering the invariant function $ F^{(d)}(p^{2},(p+q)^{2}) $ is
encoded in the nonlocal matrix element $ \left\langle \gamma \left( q\right) \left|
\overline{q}\left( x\right) \sigma _{\alpha \beta }q\left( 0\right) \right|
0\right\rangle  $, only one surviving after contracting the $ b $ quark operators.
This answers very well to the actual physical picture in $ B_{q}^{*}\rightarrow
B_{q}\gamma  $ since such a matrix element stands typically for an effect of the
magnetic interaction.

Light-cone expansion of $ \left\langle \gamma \left( q\right) \left|
\overline{q}\left( x\right) \sigma _{\alpha \beta }q\left( 0\right) \right|
0\right\rangle  $ involves, to next-to-leading order in $ x^{2} $, the leading twist 2
wavefunction $ \varphi \left( u\right)  $ and the two twist 4 distributions $
f_{1}\left( u\right)  $ and $ f_{2}\left( u\right)  $. The explicit form is
\begin{eqnarray}
\left\langle \gamma \left( q\right) \left| \overline{q}\left( x\right) \sigma _{\alpha \beta }q\left( 0\right) \right| 0\right\rangle  & = & Q_{q}\left\langle \overline{q}q\right\rangle \left( \int _{0}^{1}du\chi \varphi \left( u\right) F_{\alpha \beta }\left( ux\right) +\int _{0}^{1}dux^{2}f_{1}\left( u\right) F_{\alpha \beta }\left( ux\right) \right. \nonumber \\
 & + & \left. \int ^{1}_{0}duf_{2}\left( u\right) \left[ x_{\beta }x^{\gamma }F_{\alpha \gamma }\left( ux\right) -x_{\alpha }x^{\gamma }F_{\beta \gamma }\left( ux\right) -x^{2}F_{\alpha \beta }\left( ux\right) \right] \right) ,
\end{eqnarray}
where $ \left\langle \overline{q}q\right\rangle  $ denotes the
quark condensate density, $ \chi  $ indicates the magnetic
susceptibility of the light quark and $ F_{\alpha \beta } $ is the
external electromagnetic field tensor. Considering a simple
situation $ q\rightarrow 0 $ {[}4{]} is helpful to shed light on
the physical meaning of Eq.(29). Substituting Eq.(29) into Eq.(28)
yields
\begin{eqnarray}
F^{(d)}(p^{2},(p+q)^{2}) & = & 2Q_{q}\left\langle \overline{q}q\right\rangle \left( \chi \int _{0}^{1}du\frac{\varphi (u)}{m_{b}^{2}-(p+uq)^{2}}\right. \nonumber \\
 & - & \left. \int ^{1}_{0}du\left[ f_{1}\left( u\right) -f_{2}\left( u\right) \right] \left[ \frac{4}{\left[ m_{b}^{2}-\left( p+uq\right) ^{2}\right] ^{2}}+\frac{8m_{b}^{2}}{\left[ m_{b}^{2}-\left( p+uq\right) ^{2}\right] ^{3}}\right] \right) .
\end{eqnarray}
An emphasis we would put is that the contribution of the high
Fock-state $ q\overline{q}g $, which should be taken into account
to the present accuracy, has been neglected for it is expected to
be negligibly small from most previous calculations. For later
continuum subtraction, we have to convert Eq.(30) into a form of
dispersion integral. We do that for the twist-2 term and remain
the twist-4 ones unaffected. The relevant spectral density is easy
to evaluate by virtue of the technique in {[}13{]} once again. The
calculation procedure is as follows. First of all, we apply a
double Borel operator in $ p^{2} $ and $ \left( p+q\right) ^{2} $
to the twist-2 term, obtaining,
\begin{equation}
\widehat{B}\left( M^{2}_{1},Q_{1}^{2}\right) \widehat{B}\left( M^{2}_{2},Q_{2}^{2}\right) \int _{0}^{1}du\frac{\varphi (u)}{m_{b}^{2}-(p+uq)^{2}}=\frac{1}{M_{1}^{2}+M_{2}^{2}}e^{-\frac{M_{1}^{2}+M_{2}^{2}}{M_{1}^{2}M_{2}^{2}}m_{b}^{2}}\varphi \left( \frac{M_{1}^{2}}{M_{1}^{2}+M_{2}^{2}}\right) .
\end{equation}
The symmetry of the correlator in Eq.(2) allows us to set $
M_{1}^{2}=M_{2}^{2} $ so that the wavefunction $ \varphi \left(
\frac{M_{1}^{2}}{M_{1}^{2}+M_{2}^{2}}\right) $ can take its value
at the symmetric point $ \mu _{0}=1/2 $. Considering it and making
a replacement $ M_{1}^{2}\rightarrow \frac{1}{\sigma
_{1}},M_{1}^{2}\rightarrow \frac{1}{\sigma _{1}} $ in Eq.(31), we
get
\begin{eqnarray}
\widehat{B}\left( \frac{1}{\sigma _{1}},Q_{1}^{2}\right) \widehat{B}\left( \frac{1}{\sigma _{2}},Q_{2}^{2}\right) \int _{0}^{1}du\frac{\varphi (u)}{m_{b}^{2}-(p+uq)^{2}} & = & \frac{\sigma _{1}\sigma _{2}}{\sigma _{1}+\sigma _{2}}e^{-m_{b}^{2}(\sigma _{1}+\sigma _{2})}\varphi (1/2)\nonumber \\
 & = & f\left( \sigma _{1,}\sigma _{2}\right) .
\end{eqnarray}
Finally, we take the function $ \frac{1}{\sigma _{1}\sigma _{1}}f\left( \sigma
_{1,}\sigma _{2}\right) =\overline{f}\left( \sigma _{1,}\sigma _{2}\right)  $ and
perform one more Borel transformation in the variables $ \sigma _{1} $ and $ \sigma
_{2} $. The relevant spectral density reads
\begin{eqnarray}
\rho ^{\left( tw2\right) }\left( s_{1},s_{2}\right)  & = & \frac{1}{s_{1}s_{2}}\widehat{B}\left( \frac{1}{s_{1}},\sigma _{1}\right) \widehat{B}\left( \frac{1}{s_{2}},\sigma _{2}\right) \overline{f}\left( \sigma _{1,}\sigma _{2}\right) \nonumber \\
 & = & \delta \left( s_{1}-s_{2}\right) \varphi (1/2)\Theta \left( s_{1}-m_{b}^{2}\right) \Theta \left( s_{2}-m_{b}^{2}\right) .
\end{eqnarray}
At present, it suffices to represent the twist-2 contribution in a form of dispersion
relation. We derive the following expression for $ F^{(d)}(p^{2},(p+q)^{2}) $,
\begin{eqnarray}
F^{(d)}(p^{2},(p+q)^{2}) & = & 2Q_{q}\left\langle \overline{q}q\right\rangle \left( \chi \int _{m_{b}^{2}}^{\infty }ds_{1}ds_{2}\frac{\delta \left( s_{1}-s_{2}\right) \varphi (1/2)}{\left( p-s_{1}\right) \left[ \left( p+q\right) ^{2}-s_{2}\right] }\right. \nonumber \\
 & - & \left. \int ^{1}_{0}du\left[ f_{1}\left( u\right) -f_{2}\left( u\right) \right] \left[ \frac{4}{\left[ m_{b}^{2}-\left( p+uq\right) ^{2}\right] ^{2}}+\frac{8m_{b}^{2}}{\left[ m_{b}^{2}-\left( p+uq\right) ^{2}\right] ^{3}}\right] \right) .
\end{eqnarray}
After the Borel transformation it becomes
\begin{eqnarray}
\overline{F}^{\left( d\right) }\left( M_{1}^{2},M_{2}^{2}\right)  & = & \frac{1}{M_{1}^{2}M_{2}^{2}}\left[ 2Q_{q}\left\langle \overline{q}q\right\rangle \varphi \left( 1/2\right) \chi \int _{m_{b}^{2}}^{\infty }dse^{-\frac{s}{M^{2}}}-8Q_{q}\left\langle \overline{q}q\right\rangle \right. \nonumber \\
 & \times  & \left. \left[ f_{1}\left( 1/2\right) -f_{2}\left( 1/2\right) \right] \left( 1+\frac{m_{b}^{2}}{M^{2}}\right) e^{-\frac{m_{b}^{2}}{M^{2}}}\right] .
\end{eqnarray}

Now we conclude this section with writing down the light-cone OPE result for the
invariant function $ F\left( p^{2},\left( p+q\right) ^{2}\right)  $ in the Borel
transformed form,
\begin{equation}
\overline{F}^{\left( OPE\right) }\left( M_{1}^{2},M_{2}^{2}\right) =\overline{F}^{\left( a+b\right) }\left( M_{1}^{2},M_{2}^{2}\right) +\overline{F}^{\left( c\right) }\left( M_{1}^{2},M_{2}^{2}\right) +\overline{F}^{\left( d\right) }\left( M_{1}^{2},M_{2}^{2}\right) .
\end{equation}

\begin{center}{\large \bf 4. SUM RULES AND NUMERICAL DISCUSSIONS}\end{center}

Matching Eq.(36) with the corresponding hadronic representation
(6) and using the quark-hadron duality ansatz, in which the
hadronic spectral density $ \rho ^{H}\left( s_{1},s_{2}\right)  $
is usually assumed to coincide with that derived in the light-cone
OPE,
we arrive at the sum rule for the product
$f_{B_{q}}f_{B^{*}_{q}}g_{B^{*}_{q}B_{q}\gamma }$,
\begin{eqnarray}
f_{B_{q}}f_{B^{*}_{q}}g_{B^{*}_{q}B_{q}\gamma } & = & \frac{m_{b}+m_{q}}{m^{2}_{B_{q}}m_{B_{q}^{*}}}e^{\frac{m_{B_{q}}^{2}+m_{B_{q}^{*}}^{2}}{2M^{2}}}\left\{ 2Q_{q}\left\langle \overline{q}q\right\rangle \chi \varphi \left( 1/2\right) \left( e^{-\frac{m_{b}^{2}}{M^{2}}}-e^{-\frac{s_{0}}{M^{2}}}\right) M^{2}-2Q_{b}\left\langle \overline{q}q\right\rangle e^{-\frac{\rho m_{b}^{2}}{\rho M^{2}-1}}\right. \nonumber \\
 & - & 8Q_{q}\left\langle \overline{q}q\right\rangle \left[ f_{1}\left( 1/2\right) -f_{2}\left( 1/2\right) \right] \left( 1+\frac{m_{b}^{2}}{M^{2}}\right) e^{-\frac{m_{b}^{2}}{M^{2}}}+\frac{3m_{q}}{2\pi ^{2}}\int _{\left( m_{b}+m_{q}\right) ^{2}}^{s_{0}}dse^{-\frac{s}{M^{2}}}\nonumber \\
 & \times  & \left. \left[ \left( Q_{b}-Q_{q}\right) \lambda \left( 1,m_{b}^{2}/s,m_{q}^{2}/s\right) +Q_{q}\ln \frac{1-m_{b}^{2}/s+m_{q}^{2}/s+\lambda \left( 1,m_{b}^{2}/s,m_{q}^{2}/s\right) }{1-m_{b}^{2}/s+m_{q}^{2}/s-\lambda \left( 1,m_{b}^{2}/s,m_{q}^{2}/s\right) }\right] \right\} .\nonumber \\
 &  & \, \,
\end{eqnarray}

At this point, we have a minor comment. Nonperturbative QCD
dynamics dominates the sum rule; especially the magnetic
interaction term makes the leading contribution, while the hard
quark contribution modifies merely the sum rule at the level of $
m_{q} $, a negligibly small effect. The nonlocal matrix element $
\left\langle \gamma \left( q\right) \left| \overline{q}\left(
x\right) \gamma _{\mu }\gamma _{5}q\left( 0\right) \right|
0\right\rangle $, which starts with twist-3, is effectively
eliminated, making the sum rule free of the contamination by its
uncertainty. In addition, what we should emphasize is that our
subtraction procedure is different from that in Ref.[10], in which
continuum subtraction $e^{-\frac{m_b^2}{M^2}}\rightarrow
e^{-\frac{m_b^2}{M^2}}-e^{-\frac{s_0}{M^2}}$ is made for the
twist-3 and -4 parts.

The actual calculation needs a dynamical input. The
nonperturbative parameters concerning the light quarks include the
light-cone wavefunctions $ \varphi \left( 1/2\right)  $, $
f_{1}\left( 1/2\right)  $ and $ f_{2}\left( 1/2\right)  $, quark
condensate $ \left\langle \overline{q}q\right\rangle $, magnetic
susceptibility $ \chi  $ and light quark mass $ m_{q} $, Among
these, the leading twist-2 wavefunction $ \varphi \left(
1/2\right)  $ embodies the underlying dynamics and hence demands a
precision determination. Fortunately, it follows from a well-found
calculation that $ \varphi \left( u\right)  $ is dependent quite
weakly on the scale and deviates little from its asymptotic form
{[}4{]}, with a well-tried universality. Therefore, we can use
\begin{equation}
\varphi \left( u\right) =6u\left( 1-u\right) ,
\end{equation}
to high accuracy. According to the analysis given in {[}7{]}, the set of the
twist-4 wavefunctions can be specified as
\begin{equation}
f_{1}\left(u\right)=-\frac{1}{8}\left(1-u\right)\left(3-u\right) ,
\end{equation}
\begin{equation}
f_{2}\left( u\right)=-\frac{1}{4}\left( 1-u\right) ^{2}.
\end{equation}
Since we work in the case that the QCD radiative correlations are
neglected, the appropriate scale, at which the quark condensate
$\langle\bar{q}q\rangle$ and magnetic susceptibility $\chi$ take
values, should be set by the typical virtuality of the underlying
$b$ quark $\mu_{b}\sim \surd\overline{m_{B_q}^2-m^2_b}$, as in
Ref.[8]. We take the results used widely $ \left\langle
\overline{u}u\right\rangle (\mu =1 \ GeV)=\left\langle
\overline{d}d\right\rangle \left( \mu =1\ GeV\right) =-\left( 0.24
\ GeV\right) ^{3}, $ $ \left\langle \overline{s}s\right\rangle
\left( \mu =1\ GeV\right) =0.8\left\langle
\overline{u}u\right\rangle (\mu =1 \ GeV), $ and $ \chi (\mu =1\
GeV)=-4.4\ GeV^{-2} $ {[}15{]}, from which the use of the
Renormalization Group Equation (RGE) gets $ \left\langle
\overline{u}u\right\rangle \left( \mu _{b}\right) =\left\langle
\overline{d}d\right\rangle \left( \mu _{b}\right) =-0.018 \
GeV^{3} $, $ \left\langle \overline{s}s\right\rangle \left( \mu
_{b}\right) =-0.014\ GeV^{3} $ and $ \chi \left( \mu _{b}\right)
=-3.4\ GeV^{-2} $. Furthermore, the light-quark masses are set as
$ m_{u}=m_{d}\approx 0 $, $ m_{s}=0.115\ GeV $. As far as the $ B
$ channel parameters are concerned, we use $ m_{b}=4.8\ GeV $ for
the $ b $ quark mass and $ m_{B_{u,d}}=5.279\ GeV $, $
m_{B_{u,d}^{*}}=5.325 \ GeV $, $ m_{B_{s}}=5.369\ GeV $ and $
m_{B^{*}_{s}}=5.416\ GeV $ for the set of the $ B $ meson masses.
The relevant decay constants, for consistency, require a
recalculation in the two point sum rules in which a chiral
correlator should be chosen in a proper way. Moreover, we would do
that at the leading order in $ \alpha _{s}\left( Q^{2}\right)  $
since in present case the sum rule for $
f_{B_{q}}f_{B^{*}_{q}}g_{B^{*}_{q}B_{q}\gamma } $ works at the
same accuracy. In the light of the prescription in {[}11{]}, the
best fits lead to $f_{B_{u,d}}=120\ MeV,\, f_{B^{*}_{u,d}}=141\
MeV$ with $ s_{0}=32\ GeV^{2} $, and $f_{B_{s}}=147\ MeV,\,
f_{B^{*}_{s}}=160\ MeV$ with $ s_{0}=34\ GeV^{2}$. All the above
parameters will be used as the central values in the following
numerical estimates.

To proceed, one has to find the reasonable ranges of $ M^{2} $,
from which the desired sum rules can be read off. This is a
critical step towards deriving a reliable prediction. For this
end, we use the standard procedure where the high state
contributions are limited to a level below $ 30\% $ while the
twist-4 ones to an order less than $ 10\% $; simultaneously it is
required that the resulting sum rules be considerably stable. In
the first place, we focus on the derivation of the sum rule for $
g_{B^{*}_{u}B_{u}\gamma } $. In this case, using the above
criterion the fiducial values of $ M^{2} $are found to lie between
$ 8-14\ GeV^{2} $, with $ s_{0}=32\ GeV^{2} $. In this range,  $
M^{2} $ dependence of the product $
f_{B_{u}}f_{B^{*}_{u}}g_{B^{*}_{u}B_{u}\gamma } $, as shown in
Fig. 2, is quite weak. From the \char`\"{}window\char`\"{} we have
the sum rule results
\begin{equation}
f_{B_{u}}f_{B^{*}_{u}}g_{B^{*}_{u}B_{u}\gamma }=10.12\pm 0.17\
MeV,\, g_{B^{*}_{u}B_{u}\gamma }=0.59\pm 0.0 1\ GeV^{-1},
\end{equation}
the quoted errors being due to the change of $M^2$. Furthermore,
it is necessary to look into the uncertainties arising from the
input parameters. Let $s_{0}$ change between $ 31-33\ GeV^{2}$
while the other parameters keep fixed, the resulting variation of
the product $ f_{B_{u}}f_{B^{*}_{u}}g_{B^{*}_{u}B_{u}\gamma } $ is
roughly $ \pm 8\% $. Utilizing this result and incorporating the
corresponding changes of the decay constants, the uncertainty in
the coupling $ g_{B^{*}_{u}B_{u}\gamma } $ is estimated at the
level of $\pm 5\%$. To look at the impact of the uncertainty in $
m_{b} $, we concentrate on the sum rule for $
g_{B^{*}_{u}B_{u}\gamma } $ and consider a correlated variation
effect, which appears as $ m_{b} $ changes from $ 4.6\ GeV $ to $
5.0\ GeV $, in such a way in which we substitute the analytical
forms of the sum rules for the decay constants into Eq.(37),
observe the variation of $ g_{B^{*}_{u}B_{u}\gamma } $ with $
m_{b} $ by requiring that all the sum rules take only their values
from the best fits in both $ s_{0} $ and $ M^{2} $. A detailed
analysis shows that in this case the resulting influence amounts
to $ \pm 6\% $. We add also a numerical uncertainty of $30\%$ to
all the condensate parameters and see the effects. This results in
an error of $\pm8\%$ in the sum rule for
$g_{B^{*}_{u}B_{u}\gamma}$. Finally, the resulting total
uncertainty in $g_{B^{\ast}_{\mu}B_{\mu}\gamma}$ can be fixed at
about $21\%$, by adding linearly up all the considered errors.
Also, the same procedure is applied to the case of
$B_{d}^{*}B_{d}\gamma $; however, we would like to consider an
additional source of uncertainty in the $B_{s}^{*}B_{s}\gamma$
case, allowing $m_s$ to change in the region of $115\pm25 MeV$.
The resulting two sum rules turn out to have the Borel intervals
and uncertainties different slightly from the corresponding those
in the sum rule for $
f_{B_{u}}f_{B^{*}_{u}}g_{B^{*}_{u}B_{u}\gamma } $. We don't give
details any more for brevity. The final results read:
\begin{equation}
f_{B_{d}}f_{B^{*}_{d}}g_{B^{*}_{d}B_{d}\gamma }=-5.31\pm 0.15\ MeV,\, g_{B^{*}_{d}B_{d}\gamma }=-0.31\pm
0.01\ GeV^{-1},
\end{equation}
with $ s_{0}=32\ GeV^{2} $ and $ M^{2}=7-14\ GeV^{2} $, and
\begin{equation}
f_{B_{s}}f_{B^{*}_{s}}g_{B^{*}_{s}B_{s}\gamma }=-7.46\pm 0.23\
MeV,\, g_{B^{*}_{s}B_{s}\gamma }=-0.32\pm 0.01\ GeV^{-1},
\end{equation}
with $ s_{0}=34\ GeV^{2} $ and $ M^{2}=8-15\ GeV^{2} $. The total
uncertainties in the two coupling constants are respectively about
$ 20\%$ and $23\%$. The sum rule stability is illustrated in Fig.2
too, for the products
$f_{B_{d}}f_{B^{*}_{d}}g_{B^{*}_{d}B_{d}\gamma } $ and $
f_{B_{s}}f_{B^{*}_{s}}g_{B^{*}_{s}B_{s}\gamma } $.

Having the sum rule results for $ g_{B^{*}_{q}B_{q}\gamma} $ at
hand, we can calculate the decay widths by means of the formula
\begin{equation}
\Gamma \left( B_{q}^{*}\rightarrow B_{q}\gamma \right) =\frac{g^{2}_{B^{*}_{q}B_{q}\gamma }}{96\pi }\left( \frac{m^{2}_{B^{*}_{q}}-m^{2}_{B_{q}}}{m_{B^{*}_{q}}}\right) ^{3}.
\end{equation}
The results are
\begin{equation}
\Gamma(B_{u}^{*}\rightarrow B_{u}\gamma)=0.89\pm 0.34\ KeV,
\end{equation}
\begin{equation}
\Gamma(B_{d}^{*}\rightarrow B_{d}\gamma)=0.25\pm 0.10\ KeV,
\end{equation}
\begin{equation}
\Gamma(B_{s}^{*}\rightarrow B_{s}\gamma)=0.28\pm 0.13\ KeV.
\end{equation}

It is obvious that the sum rule results (45) and (46), within the
errors, are in agreement with the estimates from the standard
light-cone QCD sum rules {[}10{]}: $ \Gamma
\left(B_{u}^{*}\rightarrow B_{u}\gamma \right) =0.63 \ KeV $ and $
\Gamma \left( B_{d}^{*}\rightarrow B_{d}\gamma \right) =0.16\
KeV$. If we make a comparison of our sum rule predictions and
chiral perturbation theory ones {[}12{]}, on the other hand, it is
demonstrated that there is a considerable numerical difference in
the obtained decay widths in the case of $B_{u}^{*}\rightarrow
B_{u}\gamma$; but the obtained predictions are comparable with
each other within errors in the case of $B_{d}^{*}\rightarrow
B_{d}\gamma $. Introducing the ratio $ R=\Gamma \left(
B_{s}^{*}\rightarrow B_{s}\gamma \right) /\Gamma \left(
B_{d}^{*}\rightarrow B_{d}\gamma \right)$ to see $ SU(3) $
breaking effect in the radiative decays, we find $R=0.43-2.72$,
with a large uncertainty. However, the resulting lower limit of
the ratio is close to $R=1/3\approx 0.33 $ obtained in chiral
perturbation theory{[}12{]}.

\begin{center}{\large\bf 5. SUMMARY}
\end{center}

To sum up, in this paper we present an improved light-cone QCD sum
rule approach to the radiative decay $ B_{q}^{*}\rightarrow
B_{q}\gamma $ whose model estimates are at issue, an adequate
chiral current correlator being used to get a sum rule result
consistent with the underlying physics. In this framework, a
detailed derivation is given of the sum rules for the relevant
coupling constant $ g_{B^{*}_{q}B_{q}\gamma }$, which parametrizes
all the long distance dynamics. The resulting sum rules are of the
two main features: (i). It is the magnetic interaction which makes
the leading contribution to the resulting sum rules; on the
contrary, the perturbative hard quarks modify the sum rules only
at the level of $ m_{q} $. This typically is the principal
dynamical feature the radiative decay $ B_{q}^{*}\rightarrow
B_{q}\gamma  $ possesses. (ii). The nonlocal matrix element $
\left\langle \gamma \left( q\right) \left| \overline{q}\left(
x\right) \gamma _{\mu }\gamma _{5}q\left( 0\right) \right|
0\right\rangle  $, which is equally important but poorly known in
comparison with the leading matrix element $ \left\langle \gamma
\left( q\right) \left| \overline{q}\left( x\right) \sigma _{\alpha
\beta }q\left( 0\right) \right| 0\right\rangle  $ {[}10{]},
disappears in our sum rules, which controls effectively the
pollution due to the long distance effects. In addition, we apply
the light cone OPE instead of the short distance expansion to the
quark condensate contribution, minimizing the uncertainties in the
sum rules to the accuracy in consideration and guaranteeing a
self-consistent sum rule calculation. It is predicted that $
\Gamma \left( B_{u}^{*}\rightarrow B_{u}\gamma \right)
=0.89\pm0.34\ KeV,\, \Gamma \left( B_{d}^{*}\rightarrow
B_{d}\gamma \right) =0.25 \pm0.10\ KeV $ and $ \Gamma \left(
B_{s}^{*}\rightarrow B_{s}\gamma \right) =0.28\pm0.13\ KeV. $ All
these channels are predicted to be of the decay widths at the
level of $ 10^{-7}\ GeV$ and therefore are accessible at the
current running B factories and the future LHC. Also, a comparison
is made with other model predictions. There is an approximate
agreement between our results and those from the standard
light-cone sum rules {[}10{]}, for $ \Gamma \left(
B_{u}^{*}\rightarrow B_{u}\gamma \right)  $ and $ \Gamma \left(
B_{d}^{*}\rightarrow B_{d}\gamma \right)  $. We find also that the
resulting sum rule predictions deviate more or less from those of
chiral perturbation theory {[}12{]}, depending on different cases.
Striving for perfection of phenomenological models and enhancing
ones' ability in controlling nonperturbative dynamics, we believe
the procedure presented here to be a valuable attempt in the
direction.

Z. H. Li thanks Prof. K. T. Chao for helpful discussions and comments. This
work is in part supported by the National Science Foundation of China.

\newpage
\begin{figure}
\centerline{\epsfxsize=12cm \epsfbox{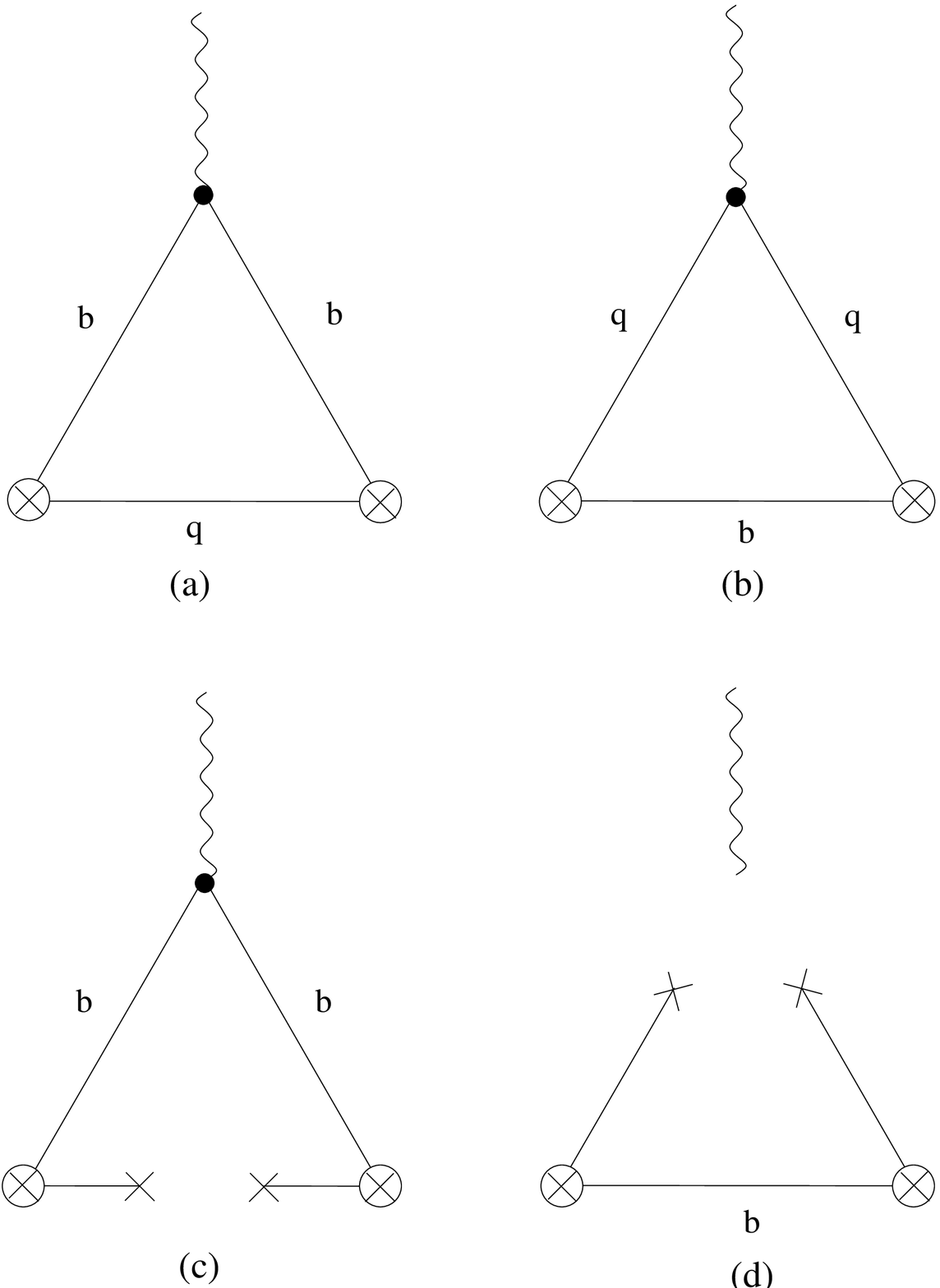}}
 \vspu \noindent
Fig.1. Diagrams making the predominate contributions to the
correlator (2). (a) and (b) depict perturbative contributions, (c)
nonlocal quark condensate correction and (d) emission of photon at
long distances parametrized by photon wavefunctions. Solid lines
denote quarks, crosses vacuum quark condensates and wavy line
photon.
\end{figure}
\newpage

\begin{figure}
\centerline {\epsfxsize=14cm \epsfbox{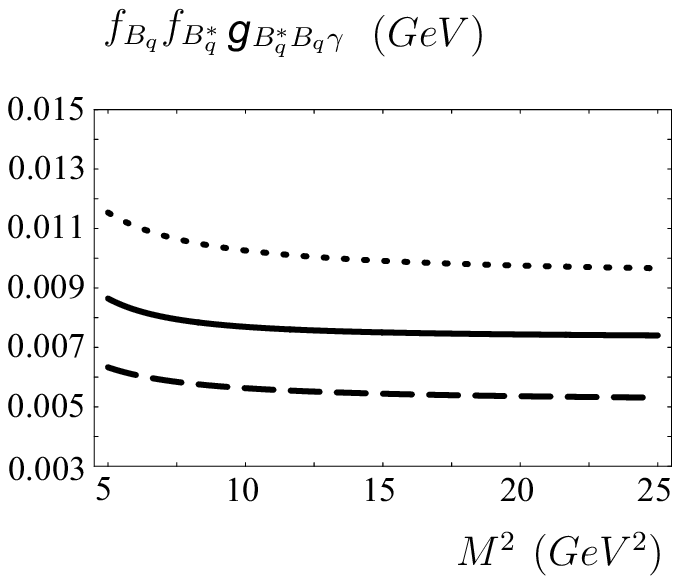}} \ \noindent
Fig.2. The light-cone QCD sum rules for $
f_{B_{q}}f_{B^{*}_{q}}g_{B^{*}_{q}B_{q}\gamma } $, with $s_{0}=32\
GeV^{2} $ for $ f_{B_{u}}f_{B^{*}_{u}}g_{B^{*}_{u}B_{u}\gamma } $
(dotted) and $ f_{B_{d}}f_{B^{*}_{d}}g_{B^{*}_{d}B_{d}\gamma } $
(dashed), and $ s_{0}=34\ GeV^{2} $ for $
f_{B_{s}}f_{B^{*}_{s}}g_{B^{*}_{s}B_{s}\gamma } $ (solid). It
should be understood that we have set $
f_{B_{d}}f_{B^{*}_{d}}g_{B^{*}_{d}B_{d}\gamma } $ and $
f_{B_{s}}f_{B^{*}_{s}}g_{B^{*}_{s}B_{s}\gamma } $ positive, for a
comparison.
\end{figure}

\newpage

\begin{thebibliography}{10}
\bibitem[1]{1}M. Beneke, G. Buchalla, M. Neubert and C. T. Sachrajda, Phys. Rev. Lett. 83
(1999) 1914; Nucl. Phys. B 591 (2000) 313.
\bibitem[2]{2}M. A. Shifman, A. I. Vainshtein, V. I. Zakharov, Nucl. Phys. B 147 (1979) 385;
448.
\bibitem[3]{3}V. L. Chernyak and I. R. Zhitnitsky, Nucl. Phys. B 345 (1990) 137.
\bibitem[4]{4}I. I. Balitsky, V. M. Braun and A. V. Kolesnichenko, Nucl. Phys. B 312 (1989)
509.
\bibitem[5]{5}V. M. Belyaev, A. Khodjamirian and R. R\"{u}ckl, Z. Phys. C 60 (1993) 349; V.
M. Belyaev, V. M. Braun, A. Khodjamirian, Phys. Rev, D 51 (1995) 6177; A. Khodjamirian,
R. R\"{u}ckl, S. Weinziel and O. Yakovlev, Phys. Lett. B 410 (1997) 275; E.
Bagan, P. Ball and V. M. Braun, Phys. Lett. B 417 (1998) 154; A. Ali, V. M.
Braun and H. Simma, Z. Phys. C 63 (1994) 437.
\bibitem[6]{6}V. L. Chernyak and A. R. Zhitnitsky, Phys. Rep. 112 (1984) 173; V. M. Braun
and I. E. Filyanov, Z. Phys. C 44 (1989) 157; C 48 (1990) 239; T. Huang, B.
Q. Ma and Q. X. Shen, Phys. Rev. D 49 (1994) 1490; P. Ball and V. M. Braun,
Phys. Rev. D 54 (1996) 2182; D 55 (1997) 5561.
\bibitem[7]{7}A. Ali and V. M. Braun, Phys. Lett. B 359 (1995) 223.
\bibitem[8]{8}A. Khodjamirian, G. Stoll and D. Wyler, Phys. Lett. B 358 (1995) 129.
\bibitem[9]{9}G. Eliam, I. Helperin and R. Mendel, Phys. Lett. B 361 (1995) 137.
\bibitem[10]{10}T. M. Aliev, D. A. Demir, E. Iltan and N. K. Pak, Phys. Rev. D 54 (1996) 857.
\bibitem[11]{11}T. Huang and Z. H. Li, Phys. Rev. D 57 (1998) 1993; T. Huang, Z. H. Li and H.
D. Zhang, J. Phys. G 25 (1999) 1179; T. Huang, Z. H. Li and X. Y. Wu, Phys.
Rev. D 63 (2001) 094001.
\bibitem[12]{12}J. F. Amundson, C, G. Boyd, I. Jenkins, M. Luke, A, V. Manohar, J. L. Rosner,
M. J. Savage and M. B. Wise, Phys. Lett. B 296 (1992) 415.
\bibitem[13]{13}V. A. Nesterenko and A. V. Radyushkin, Sov. J. Nucl. Phys. 39 (1984) 811; V.
A. Beylin and A. V. Radyushkin, Nucl. Phys. B 260 (1985) 61.
\bibitem[14]{14}E. V. Shuryak, Nucl. Phys. B 328 (1989) 85; S. V. Mikhailov and A. V. Radyushkin,
Phys. Rev. D 45 (1992) 1754; Sov. J. Nucl. Phys. 49 (1989) 494; P. Ball, V.
M. Braun and H. G. Dosch, Phys, Lett. B 273 (1991) 316.
\bibitem[15]{15}V. M. Belyaev and Ya. I. Kogan, Sov. J. Nucl. Phys. 40 (1984) 659; I. I. Balitsky,
A. V. Kolesnichenko and A. V. Yang, Sov. J. Nucl. Phys. 41 (1985) 178.
\end{thebibliography}
\end{document}